\documentclass[twocolumn]{aastex631}

\usepackage{amsmath}
\usepackage{ragged2e}
\usepackage{tabularx}

\newcommand{\prot}{\ensuremath{P_{\mbox{\scriptsize rot}}}}
\newcommand{\bprp}{\ensuremath{G_{\rm BP} - G_{\rm RP}}}

\newcommand{\ro}{$Ro$}

\newcommand{\teff}{$T_{\rm eff}$}
\newcommand{\ith}{\ensuremath{^{\rm th}}}
\newcommand{\logprot}{log$_{10}(P_{\rm rot})$}
\newcolumntype{P}[1]{>{\centering\arraybackslash}p{#1}}

\begin{document}

\title{In this Day and Age: An Empirical Gyrochronology Relation for Partially and Fully Convective Single Field Stars}

\correspondingauthor{Yuxi(Lucy) Lu}
\email{lucylulu12311@gmail.com}

\newcommand{\amnh}{American Museum of Natural History, Central Park West, Manhattan, NY, USA}
\newcommand{\cca}{Center for Computational Astrophysics, Flatiron Institute, 162 5\ith\ Avenue, Manhattan, NY, USA}
\newcommand{\columbia}{Department of Astronomy, Columbia University, 550 West 120\ith\ Street, New York, NY, USA}

\author[0000-0003-4769-3273]{Yuxi(Lucy) Lu}
\affiliation{\amnh}

\author[0000-0003-4540-5661]{Ruth Angus}
\affiliation{\amnh}
\affiliation{\cca}

\author[0000-0002-9328-5652]{Daniel Foreman-Mackey}
\affiliation{\cca}

\author[0000-0002-0842-863X]{Soichiro Hattori}
\affiliation{\columbia}



\begin{abstract}
Gyrochronology, the field of age-dating stars using mainly their rotation periods and masses, is ideal for inferring the ages of individual main-sequence stars. 
However, due to the lack of physical understanding of the complex magnetic fields in stars, gyrochronology relies heavily on empirical calibrations that require consistent and reliable stellar age measurements across a wide range of periods and masses. 
In this paper, we obtain a sample of consistent ages using the gyro-kinematic age-dating method, a technique to calculate the kinematics ages of stars.
Using a Gaussian Process model conditioned on ages from this sample ($\sim$ 1 - 14 Gyr) and known clusters (0.67 - 3.8 Gyr), we calibrate the first empirical gyrochronology relation that is capable of inferring ages for single, main-sequence stars between 0.67 Gyr to 14 Gyr.
Cross-validating and testing results suggest our model can infer cluster and asteroseismic ages with an average uncertainty of just over 1 Gyr. 
With this model, we obtain gyrochronology ages for $\sim$ 100,000 stars within 1.5 kpc of the Sun with period measurements from Kepler and ZTF, and 384 unique planet host stars. 
\end{abstract}

\keywords{Stellar ages -- Stellar rotation -- Catalogs -- Gaussian Processes regression -- Main sequence stars}


\section{Introduction} \label{sec:intro}
Gyrochronology \citep{Barnes2003} is a method to age-date stars mainly using their rotation periods (\prot) and mass/temperature (\teff) measurements.
It is based on the principle that stars lose angular momentum through magnetized winds and therefore, spin down with time \citep{Kraft1967}.
The simplest form of Gyrochronology relation is discovered by \cite{Skumanich1972}, stating that \prot $\propto$ ${\rm Age}^{1/2}$.

Unfortunately, This simple picture is heavily challenged by the emergence of large photometric surveys in the recent decade such as Kepler \citep{kepler}, K2 \citep{K2}, TESS \citep{TESS}, MEarth \citep{MEarth}, and ZTF \citep{ztfdata, ztftime}.  
These photometric surveys provided valuable data to measure stellar rotation in mass quantities
\citep[e.g.][]{McQuillan2013, McQuillan2014, Garcia2014, Santos2019, Santos2021, Gordon2021, Lu2022, Holcomb2022, Claytor2023}. 
These catalogs show sub-structures in the density distribution of stars in \prot-\teff\ space, suggesting not all stars spin down ``Skumanich style''.
Some of the discoveries include: the upper boundary or pile-up of solar-like stars with intermediate ages \citep{Angus2015, Hall2021, David2022} that could be caused by weakened magnetic braking \citep[e.g.][]{vansaders2016, Metcalfe2022} or perhaps the transition of latitudinal differential rotation \citep{Tokuno2022}; the intermediate period gap in partially convective GKM dwarfs \citep{McQuillan2013, Gordon2021, Lu2022} most likely caused by stalled spin-down of low-mass stars \citep{Curtis2020, Spada2020}; the bi-modality of fast and slow-rotating M dwarfs that is difficult to explain with traditional models of angular-momentum loss \citep{Irwin2011, Berta2012, Newton2017, Pass2022, Garraffo2018}; the abrupt change in stellar spin-down across the fully convective boundary \citep[Chiti et al. in prep.]{Lu2023}. 
Therefore, modern-day gyrochronology heavily relies on empirical calibrations with benchmark stars such as those with asteroseismic ages \citep[e.g.][]{Angus2015, Hall2021}, those in wide binaries \citep[][Chiti et al. in prep.]{Pass2022, SilvaBeyer2022, Otani2022, Gruner2023}, and open cluster members \citep[e.g.][]{Curtis2020, Agueros2018, Gaidos2023, Dungee2022, Bouma2023}.
Asteroseismic ages can be accurate and precise to the 10\% level with time series from Kepler.
Unfortunately, asteroseismic signal strength/frequency decreases/increases dramatically as the mass of a star decreases, and no signals have been detected for low-mass M dwarfs. 
Open clusters are generally young as they typically dissolve in the Milky Way on a time-scale of $\sim$ 200 Myr.
Much effort has been put into calibrating gyrochronology with wide binaries, however, no large catalog of consistent ages for wide binary stars currently exists. 
As a result, none of the above benchmark stars can provide a consistent sample of reliable ages for stars of vastly different masses and periods that can be used to calibrate empirical gyrochronology relations across a wide range of ages.

Recently, gyro-kinematic age-dating \citep{Angus2020, Lu2020}, a method to obtain kinematic ages from stars with similar \prot-\teff-$M_G$-Rossby Number (Ro; \prot\ divided by the convective turnover time), provide an opportunity to obtain a consistent benchmark sample for calibrating a fully empirical gyrochronology relation. 
One discovery using the ages obtained from this method is the fundamentally different spin-down law for fully and partially convective stars \citep{Lu2023}, as a result, it is important to obtain gyrochronology ages separately for partially and fully convective stars. 
By combining period measurements from Kepler and ZTF, we obtain gyro-kinematic ages for $\sim$ 50,000 stars and present the first fully empirical gyrochronology relation that is able to infer ages for single main-sequence stars of age 0.67-14 Gyr. 
In section~\ref{sec:method}, we describe the dataset, the method used to calibrate this gyrochronology relation, and the cross-validation test.
In section~\ref{sec:result}, we present the testing set and a catalog of $\sim$ 100,000 stars with gyrochronology ages. 
In section~\ref{sec:limit}, we discuss the limitations, including the effect of metallicity, and future improvements.

\section{Data \& Method} \label{sec:method}
\subsection{Data}\label{subsec:periods}
\subsubsection{Rotation Period (\prot), Rossby Number (\ro), Temperature (\teff), Absolute G Magnitude ($M_G$), and Radial Velocity (RV) Data}\label{subsec:P_RV_data}

We de-reddened $G_{\rm BP}-G_{\rm RP}$, $M_G$ measurements from Gaia DR3 using {\tt dustmap} \citep{Green2018, Green20182}.
The temperature is then calculated from $G_{\rm BP}-G_{\rm RP}$ using a polynomial fit taken from \cite{Curtis2020}.
\ro\ is calculated as \ro=\prot/$\tau_c$, in which $\tau_c$ is the convective turn-over time that depends only on the temperature of the star (See et al. in prep.).

\begin{figure}[hbt!]
    \centering
    \includegraphics[width=0.4\textwidth]{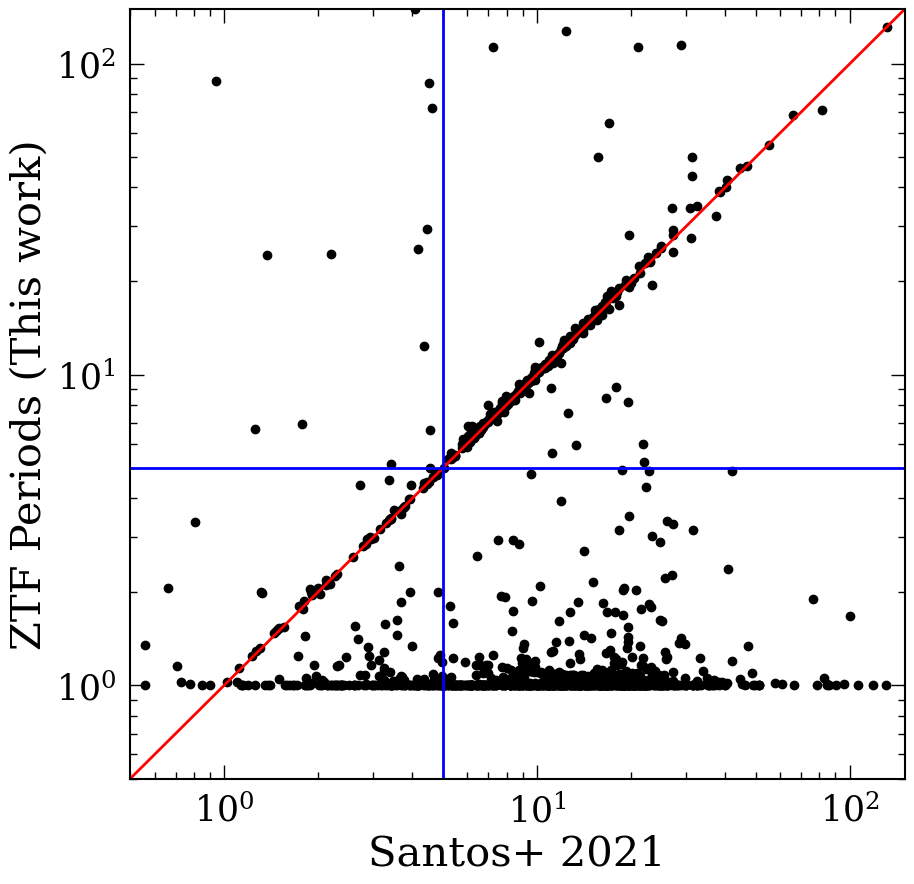}
    \caption{1-to-1 comparison between 1,270 stars with period measured from Kepler \citep{Santos2021} and ZTF \citep[][This work]{Lu2022}. 
    We found a 81\% agreement within 10\% for stars with ZTF period measurements $>$ 4 days. }
    \label{fig:figpre}
\end{figure}

\begin{figure}[hbt!]
    \centering
    \includegraphics[width=0.45\textwidth]{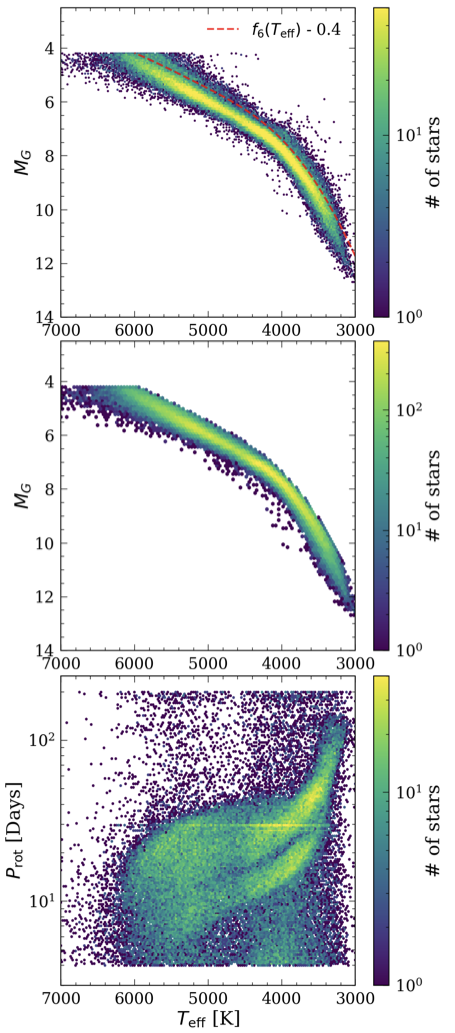}
    \caption{{\it Top}: $M_G$-\teff\ for the full sample of $\sim$ 85,000 dwarf stars with period measurements from ZTF and Kepler \citep[][this work]{McQuillan2014, Garcia2014, Santos2019, Lu2022}.
    The red dashed line shows the shifted 6\ith-order polynomial ($f_6 (T_{\rm eff})$) fitted to the entire sample that separates the equal-mass binaries from the rest of the sample.  
    {\it Middle}: similar to the top plot but after excluding equal-mass binaries (a total of 68,378 dwarf stars). 
    {\it Bottom}: period distribution of the 68,378 dwarf stars.}
    \label{fig:fig1}
\end{figure}

We obtained rotation periods for ZTF stars with Gaia G band magnitude between 13 to 18 and $G_{\rm BP}-G_{\rm RP}$ $<$ 1 using the method described in \cite{Lu2022}. 
For ZTF stars with $G_{\rm BP}-G_{\rm RP}$ $>$ 1, we adapted the rotation period measurements (before vetting) from \cite{Lu2022}.
From the full sample, we selected stars with agreeing periods from at least 2 seasons.
By comparing 1,270 overlapping period measurements from ZTF and Kepler \citep{Santos2021}, we found an 81\% agreement within 10\% for stars with ZTF period measurements $>$ 4 days (see Figure~\ref{fig:figpre}). 
As a result, we rejected stars with ZTF period measurements $<$ 4 days.
To roughly select dwarf stars, we also excluded stars with $M_G$ $<$ 4.2 mag. 
This yielded $\sim$ 55,000 ZTF stars with RV measurements from Gaia DR3 \citep{gaiadr3}. 
Combining $\sim$ 30,000 Kepler stars with $M_G$ $>$ 4.2 mag from \cite{Lu2020} with RV measurements from Gaia DR3, LAMOST \citep{lamost}, and inferred RV from \cite{Angus2022}, we obtained a total of $\sim$ 85,000 stars with RV and relatively reliable period measurements (See Figure~\ref{fig:fig1} top plot).

We then excluded equal-mass binaries by fitting a 6\ith-order polynomial ($f_6 (T_{\rm eff})$) to the entire sample and only selecting stars with $M_G > f_6 (T_{\rm eff}) - 0.4$ (shifted by eye).
We also excluded stars with \ro\ $>$ 10. 
This left us with a final sample of 68,378 stars (ZTF: 49,928; Kepler: 18,450).
The period distribution for the final sample is shown in the bottom plot of Figure~\ref{fig:fig1}.
The overall period distribution agrees with that of \cite{McQuillan2014, Santos2021}, except for an over-density at $\sim$ 4,000 K with \prot\ $<$ 10 days.
Since we did not impose conservative vetting criteria, this over-density is most likely caused by systematic.
We also see a systematic over-density at $\sim$ 30 days, this is a known systematic in ZTF, which is caused by the orbit of the moon \citep{Lu2022}.

\subsubsection{Cluster data}\label{subsec:cluster_data}
\begin{figure}[hbt!]
    \centering
    \includegraphics[width=0.45\textwidth]{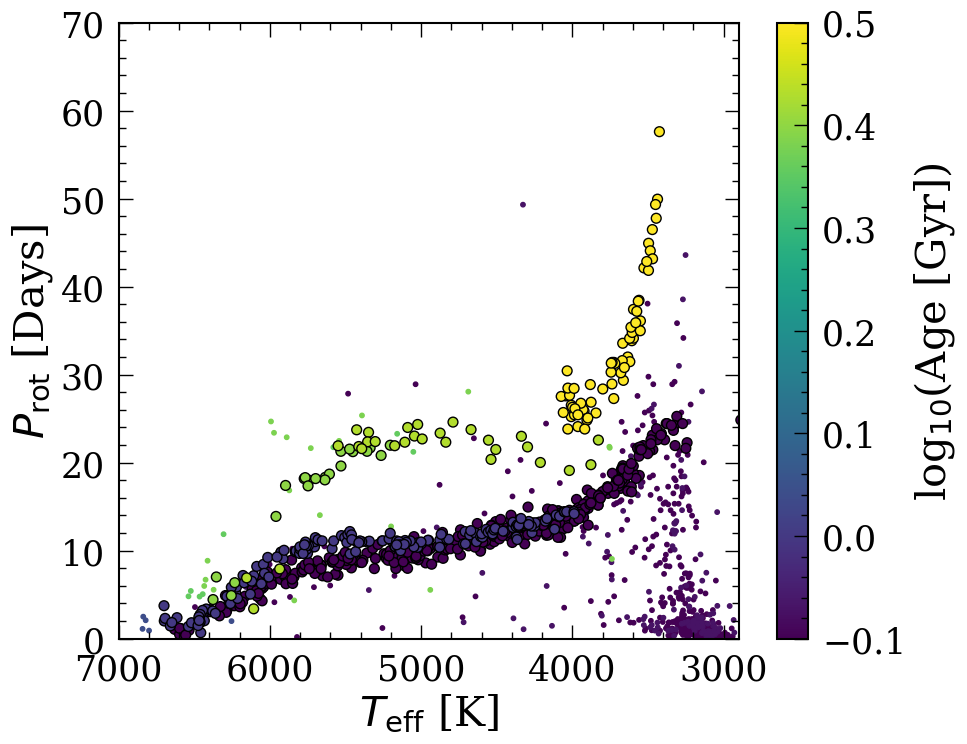}
    \caption{Full cluster sample (small points) and the final 660 cluster stars (large points) used in training. 
    The cluster ages range from 670 Myr to 4 Gyr, and the final training set is selected with 3-sigma clipping to exclude fast-rotating stars that have not converged onto the slow-rotating sequence.}
    \label{fig:fig3}
\end{figure}

Period measurements for the 4 Gyr open cluster, M67, are taken from \cite{Dungee2022}. The rest of the cluster data is taken from \citet{Curtis2020}, which includes Praesepe \citep[670 Myr;][]{Douglas2019}, Hyades \citep[730 Myr;][]{Douglas2019}, NGC~6811 \citep[1 Gyr;][]{Curtis2019}, NGC~6819 \citep[2.5 Gyr;][]{Meibom2015}, and Ruprecht~147 \citep[2.7 Gyr;][]{Curtis2020}.
We then performed a 3-sigma clipping to exclude stars that had not converged onto the slow-rotating sequence. 
The final cluster sample used in training the model included 660 stars ranging from 670 Myr to 4 Gyr (see Figure~\ref{fig:fig3}).

\subsection{Methods}
\subsubsection{Gyro-kinematic age data}
\begin{figure}[hbt!]
    \centering
    \includegraphics[width=0.48\textwidth]{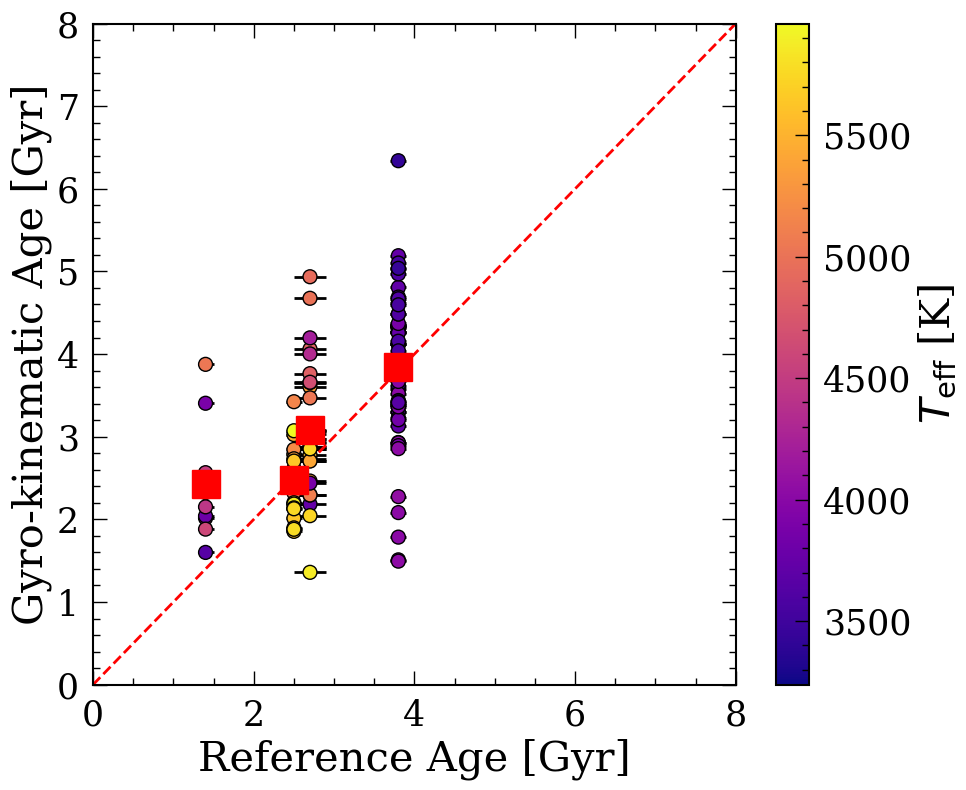}
    \caption{Optimization result comparing individual cluster ages $> 1.5$ Gyr with $M_G > 4.2$ and $R_{\rm o} < 2$ \citep{Curtis2020, Dungee2022} and gyro-kinematic ages (this work).
    The red squares show the mean gyro-kinematic ages for individual clusters.}
    \label{fig:fig2}
\end{figure}

We determined gyro-kinematic ages following the procedure described in \cite{Lu2021}, where the vertical velocity dispersion for each star is calculated from vertical velocities of stars that are similar in \prot, \teff, $M_G$, and \ro\ to the targeted star. 
We then converted the velocity dispersion measurements into stellar ages using an age-velocity-dispersion relation in \cite{Yu2018}.
The vertical velocities are calculated from Gaia DR3 proper motions \citep{gaiadr3} and RVs from various sources (data sample see Section~\ref{subsec:P_RV_data}) by transforming from the Solar system barycentric ICRS reference frame to Galactocentric Cartesian and cylindrical coordinates using {\tt astropy} \citep{astropy:2013, astropy:2018}.
The bin size to select similar stars to the targeted star in order to calculate gyro-kinematic ages was [$T_{\rm eff}$, log$_{10}(P_{\rm rot})$, $R_{\rm o}$, $M_G$] = [177.8 K, 0.15 dex, 0.15 dex, 0.2 mag].
This bin size is optimized by performing a grid search in the binning parameters ($T_{\rm eff}$, log$_{10}(P_{\rm rot})$, $R_{\rm o}$, $M_G$) and minimizing the total $\chi^2$ in predicting individual cluster ages $> 1.5$ Gyr with $M_G > 4.2$ and $R_{\rm o} < 2$ (data sample see Section~\ref{subsec:cluster_data}).
We did not use clusters with age $< 1.5$ Gyr in this process as gyro-kinematic ages for stars $< 1.5$ Gyr is heavily contaminated by binaries, and will overestimate cluster ages and produce unreliable results \citep[See Figure 4 or A.1 in][]{Lu2021}.
Figure~\ref{fig:fig2} shows the final optimization result.

We excluded stars with gyro-kinematic age $<$ 1.5 Gyr or $>$ 14 Gyr as it is possible that a significant number of the youngest stars have not yet converged onto the slow-rotating sequence, and those that are very old are likely outliers. 
The sample of 46,362 stars with corrected gyro-kinematic ages between 1.5 and 14 Gyr and cluster ages between 0.67 Gyr and 4 Gyr are shown in Figure~\ref{fig:fig4} top plot.

\subsubsection{A fully empirical gyrochronology relation with Gaussian Process}
Gaussian Processes (GP) is a generic supervised learning method designed to solve regression or classification problems.
It has been applied frequently in time-domain astronomy \citep[e.g.][]{ForemanMackey2017, Angus2018, Gilbertson2020} as it can model the covariance between the noise in the data.
Typically, a GP regressor is composed of a mean function ($m$; Equation~\ref{eq:mean}), which is ideally physically motivated, and a covariance function ($k$; Equation~\ref{eq:kern}) that captures the details that the mean function has missed. 
For a more detailed review of GP and its applications in astronomy, we direct the readers to \cite{Aigrain2022}.
In this paper, we used the \texttt{PYTHON} package \texttt{tinygp} \citep{tinygp} to construct our GP model. 

Since there is an abrupt change in the spin-down law across the fully convective boundary \citep{Lu2023}, we fitted separate GP relations to the partially and fully convective stars.
The division was made using the gap discovered in the color-magnitude-diagram (CMD).
This gap is an under-density in the CMD near the fully convective boundary and can be approximated by a line connecting [$M_G$, \bprp] $\sim$ [10.09 mag, 2.35 mag] and [$M_G \sim$, \bprp] $\sim$ [10.24 mag, 2.55 mag] \citep{Jao2018}.
It is thought to be caused by structural instabilities due to the non-equilibrium fusion of $^3$He \citep{vansanders2012, Baraffe2018, MacDonald2018, Feiden2021}.

\begin{figure}[hbt!]
    \centering
    \includegraphics[width=0.45\textwidth]{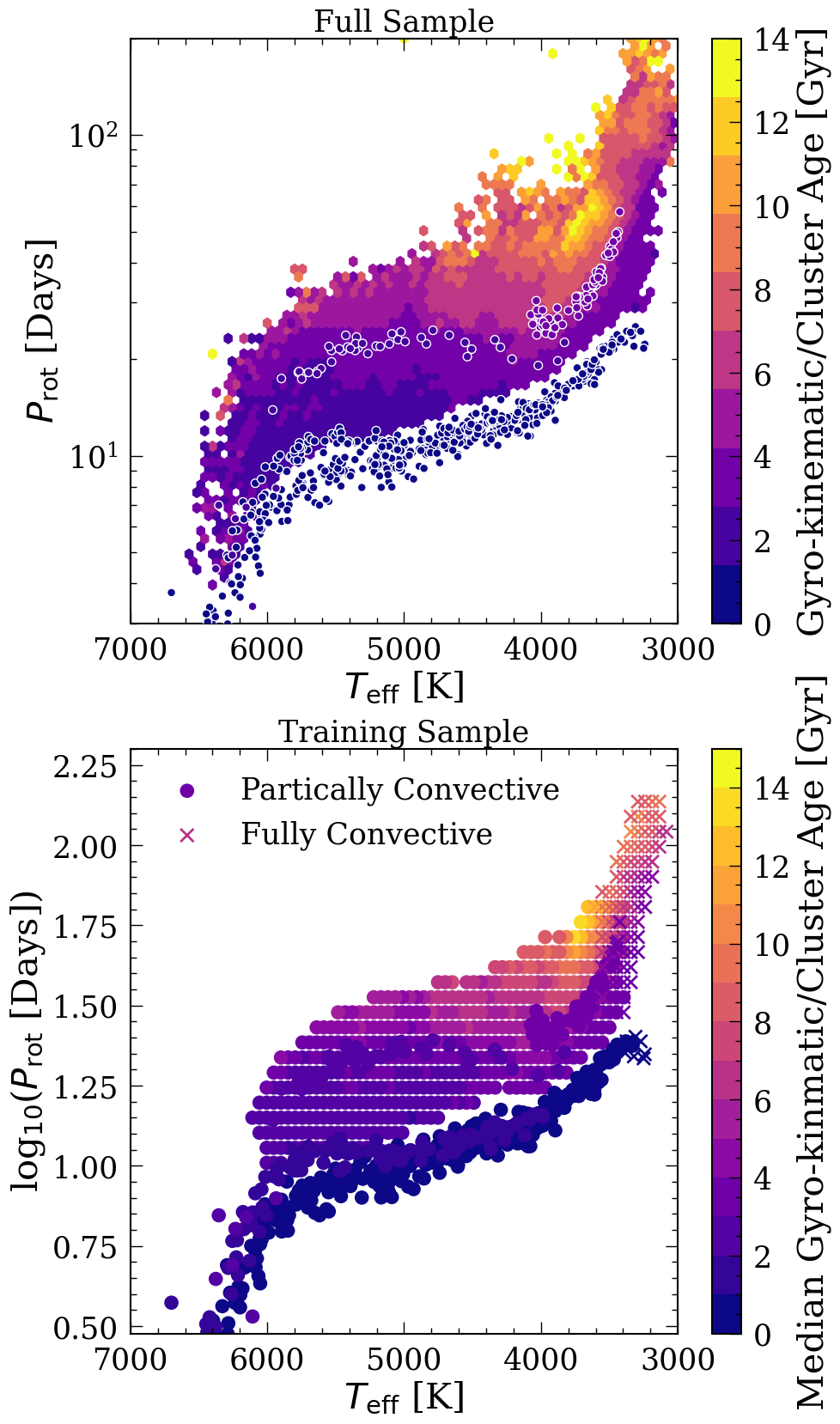}
    \caption{{\it Top}: 46,362 stars with corrected gyro-kinematic ages between 1.5 and 14 Gyr (background histogram) and individual cluster stars (circles) with literature ages between 0.67 Gyr and 4 Gyr. 
    {\it Bottom}: GP training set for the partially (circles; 1,109 data points) and fully convective (crosses; 96 data points) stars colored by the median gyro-kinematic ages in each [\teff, log$_{10}$(\prot)] $\sim$ [50 K, log$_{10}$(1 Days)] bin or the cluster ages.}
    \label{fig:fig4}
\end{figure}

As fitting a multi-dimensional GP requires a large amount of computational resources, and it is not possible to fit to all $\sim$46,000 stars with gyro-kinematic ages within a reasonable amount of time, we constructed the final training sample by dividing the stars with gyro-kinematic ages into bins with size [\teff, log$_{10}$(\prot)] $\sim$ [50 K, log$_{10}$(1 Days)] and calculating the median age in each bin if more than 10 stars were included. 
The fit was done separately for fully and partially convective stars as some of them overlap in \teff-\prot\ space.
The temperature bin size is chosen based on the estimated uncertainty in temperature measurements of $\sim$ 50K, and the period bin size is chosen so that we can obtain enough training samples for the GP. 
The uncertainties associated with the training sample are measured with the standard deviation on the gyro-kinematic ages for stars in each bin.
We then added all the individual cluster stars to the training sample and inflated their age uncertainty to be 0.5 Gyr to ensure a smooth GP fit. 
We found that using the true cluster age uncertainties reported in the literature, the GP over-fits the cluster data.
The training sample for the partially (circles; 1,109 data points) and fully convective (crosses; 96 data points) stars colored by the median gyro-kinematic or the cluster ages are shown in Figure~\ref{fig:fig4} bottom plot. 

Classical gyrochronology relations assume the age of a star can be approximated with a separable function in temperature and period, and we constructed our mean function motivated by this relation. 
We formulated the mean function to be a double broken power law in \prot\ and $T_n$ for partially convective stars to capture the sudden increase of rotation periods of M dwarfs at $\sim$3,500 K and the plateauing of the rotation periods for G/K stars at $\sim$5,000 K.
We define $T_n$ as the normalized temperature given by $T_n:=(7000-T_{\rm eff})/(7000-{\bf T^{\rm break}})$ for the partially convective stars, and $T_n:=(3500-T_{\rm eff})/500$ for the fully convective stars, in which ${\bf T^{\rm break}}$ is the temperature at which the temperature power law changes.
For fully convective stars, we used a single power law in $T_n$ and a double broken power law in rotation period, $P$ or $P_{\rm rot}$, since the temperature range for the fully convective stars is small.

In equations, The mean function is defined to be:
\begin{equation}
    m(P, T_{\rm eff}) = {\bf a}*f(P)*g(T_n)
    \label{eq:mean}
\end{equation}

where the broken power law in rotation, $f(P_{\rm rot})$, is defined to be:
\begin{equation}
    \begin{aligned}
    f(P) &= S_h^P(P, {\bf w_P})P^{\bf d_P^1}+ \\
    & S_l^P(P, {\bf w_P})P^{\bf d_P^2}{\bf P^{\rm break}}^{{\bf d_P^1}-{\bf d_P^2}}
    \end{aligned}
    \notag
\end{equation}

where ${\bf P^{\rm break}}$ is the rotation period at which the rotation power law changes. 
$S_h^P(P_{\rm rot}, {\bf w_P})$ and $S_l^P(P_{\rm rot}, {\bf w_P})$ are the smoothing functions in period space, defined to be:
\begin{equation}
    \begin{aligned}
    S_h^P&(P, {\bf w_P}) =\\
    & = \frac{1}{(1+\exp{(-(\rm log_{10}{\bf P^{\rm break}}-{\rm log_{10}P)/{\bf w_P}}))}}\\
    S_l^P&(P, {\bf w_P}) =\\
    & = \frac{1}{(1+\exp{(-(-\rm log_{10}{\bf P^{\rm break}}+{\rm log_{10}P)/{\bf w_P}}))}}\\
    \end{aligned}
    \notag
\end{equation}

The broken power law in temperature, $g(T_n)$, is defined to be:
\begin{equation}
    \begin{aligned}
    g(T_n) &= S_h^T(T_n, {\bf w_T})(T_n-{\bf c_T})^{\bf d_T^1} + \\
    & S_l^T(T_n, {\bf w_T})(T_n-{\bf c_T})^{\bf d_T^2}(1-{\bf c_T})^{{\bf d_T^1}-{\bf d_T^2}}
    \end{aligned}
    \notag
\end{equation}

where $S_h^T(T_n, {\bf w_T})$ and $S_l^T(T_n, {\bf w_T})$ are the smoothing functions in temperature space, defined to be:
\begin{equation}
    \begin{aligned}
    S_h^T(T_n, {\bf w_T}) = 1/(1+\exp{(-(1-T_n)/{\bf w_T})})\\
    S_l^T(T_n, {\bf w_T}) = 1/(1+\exp{(-(-1+T_n)/{\bf w_T})})
    \end{aligned}
    \notag
\end{equation}

The bold letters show the variables that will be fitted to the data, they are defined in Table~\ref{tab:tab1}.
The smoothing functions (e.g. $S_h^P$ and $S_h^T$) can be viewed as switches for the broken power laws.
${\bf w_P}$ and ${\bf w_T}$ dictate how smooth the broken power laws are (e.g. ${\bf w_P}$=0 or ${\bf w_T}$=0 indicate a sharp transition between the power laws).
Since the fully convective stars only span a small range in temperature, we used a single power law, so that $g(T_n) = (T_n-{\bf c_T})^{\bf d_T^1}$.

For the covariance function of the GP model, we used a 2-D uncorrelated squared exponential kernel, meaning we assume no correlation between the temperature and period measurements.
The function is defined to be:

\begin{equation}
    \begin{aligned}
    & k_{\rm SE}(T_{\rm eff}, T_{\rm eff}', \log_{10}(P_{\rm rot}), \log_{10}(P_{\rm rot})') = \\ 
    & = \sigma^2\exp{(\frac{(T_{\rm eff} - T_{\rm eff}')^2}{2{\bf l_{\rm T}}^2})}
    \exp{(\frac{(\log_{10}(P_{\rm rot}) - \log_{10}(P_{\rm rot})')^2}{2{\bf l_{\rm logP}}^2})}
    \end{aligned}
    \label{eq:kern}
\end{equation}
where $T_{\rm eff}$, $T_{\rm eff}'$ are two different data points in temperature space (same for $\log_{10}(P_{\rm rot})$ and $\log_{10}(P_{\rm rot})'$).
${\bf l_{\rm T}}$ and ${\bf l_{\rm P}}$ determine the length scale of the correlation between temperature measurements and period measurements, respectively. 
$\sigma^2$ determines the strength of the correlation.
In other words, the covariance function determines how the response at one temperature and period point is affected by responses at other temperature and period points.

\begin{figure}
    \centering
    \includegraphics[width=0.48\textwidth]{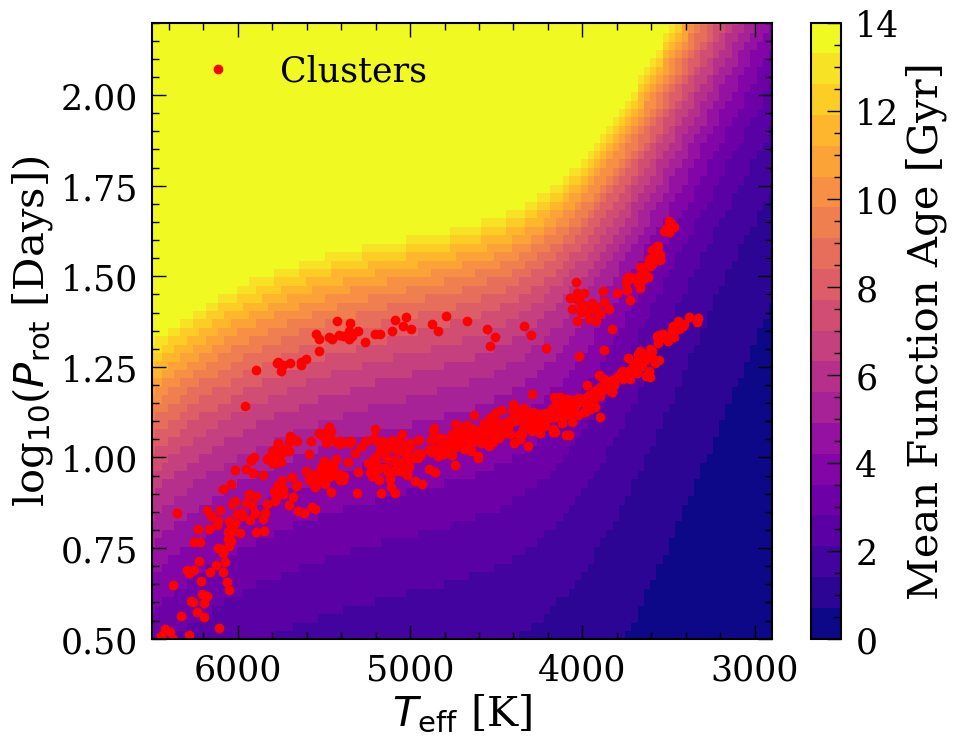}
    \caption{Age predicted from the mean function calculated using the initial values shown in Table~\ref{tab:tab1}. 
    The red points show the cluster star sample. 
    The mean function is flexible enough to capture the cluster shapes.}
    \label{fig:meanfunc}
\end{figure}

The initial values for the parameters used in the mean function and covariance function before optimizing are shown in Table~\ref{tab:tab1}.
Figure~\ref{fig:meanfunc} shows the mean function (background) calculated with the initial values and the cluster members overlayed on top (red points). 

\begin{table*}
    \begin{center}
        \caption{Initial values for maximizing the log-likelihood function, Gaussian prior width used in the MCMC fits, and and final values for the mean function (Equation~\ref{eq:mean}) and GP squared exponential kernel (Equation~\ref{eq:kern}) parameters after the MCMC fitting.}
        \begin{tabularx}{0.8\textwidth}{P{1.5cm}|P{3cm}|P{2cm}|P{2cm}|P{2cm}|P{2cm}}
            Parameters for mean function & Descriptions & Initial value & Gaussian prior width & Best-fit value (partially convective stars) & Best-fit value (fully convective stars) \\
            \hline \hline 
            a & Amplitude of the mean function & 0.3 &40 & $118.969^{36.161}_{-35.128}$  & $0.774^{0.008}_{-0.008}$ \\
            \hline
            $d_{\rm P}^1$ & Power index for stars with \prot\ $>$ P$_{\rm rot}^{\rm break}$ & 0.8& 0.2 & $1.822^{0.122}_{-0.112}$ & $1.811^{0.018}_{-0.018}$ \\
            \hline
            $d_{\rm P}^2$ & Power index for stars with \prot\ $<$ P$_{\rm rot}^{\rm break}$ & 1& 0.5 & $-0.405^{0.117}_{-0.118}$ & $0.367^{0.004}_{-0.004}$  \\
            \hline
            c & shift in the temperature scale & -0.5& 0.2 & $-0.399^{0.097}_{-0.105}$ & $-0.223^{0.002}_{-0.002}$ \\
            \hline
            $d_{\rm T}^1$ & Power index for stars with \teff\ $>$ T$^{\rm break}$ & -1& 0.5 & $1.646^{0.486}_{-0.478}$ & $-0.687^{0.007}_{-0.007}$ \\
            \hline
            $d_{\rm T}^2$ & Power index for stars with \teff\ $<$ T$^{\rm break}$ & -10& 6 & $-17.779^{3.043}_{-3.545}$ &   \\
            \hline
            P$^{\rm break}$ & \prot\ at which the period power law breaks & 30& 30 & $100.836^{21.173}_{-15.663}$ & $73.322^{0.727}_{-0.700}$ \\
            \hline
            T$^{\rm break}$ & \teff\ at which the temperature power law breaks & 4000& 500 & $3713.699^{53.318}_{-49.993}$ &   \\
            \hline
            $w_{\rm T}$ & Smoothness of the temperature power law transition & 0.1& 0.01 & $0.062^{0.008}_{-0.007}$ & \\
            \hline
            $w_{\rm P}$ & Smoothness of the period power law transition & 0.1& 0.01 & $0.111^{0.010}_{-0.010}$ & $0.068^{0.001}_{-0.001}$ \\
            \hline \hline 
            Parameters for the kernel function & Descriptions & Initial value & Gaussian prior width& Best-fit value (partially convective stars) & Best-fit value (fully convective stars) \\
            \hline \hline 
            ln($\sigma$) & log of the kernel amplitude & 0& 0.5  & $-2.070^{0.282}_{-0.260}$ & $-1.004^{0.102}_{-0.099}$ \\
            \hline
            ln(l$_{\rm T}$) & log of the scaling in temperature & 1& 1  & $5.619^{0.214}_{-0.223}$ & $6.532^{0.909}_{-0.875}$ \\
            \hline
            ln(l$_{\rm logP}$) & log of the scaling in \logprot & 1& 1  & $-2.573^{0.254}_{-0.236}$ & $-1.001^{0.098}_{-0.097}$ 
            \label{tab:tab1}
        \end{tabularx}
    \end{center}
    
\end{table*}

\begin{figure*}[htb!]
    \centering
    \includegraphics[width=0.95\textwidth]{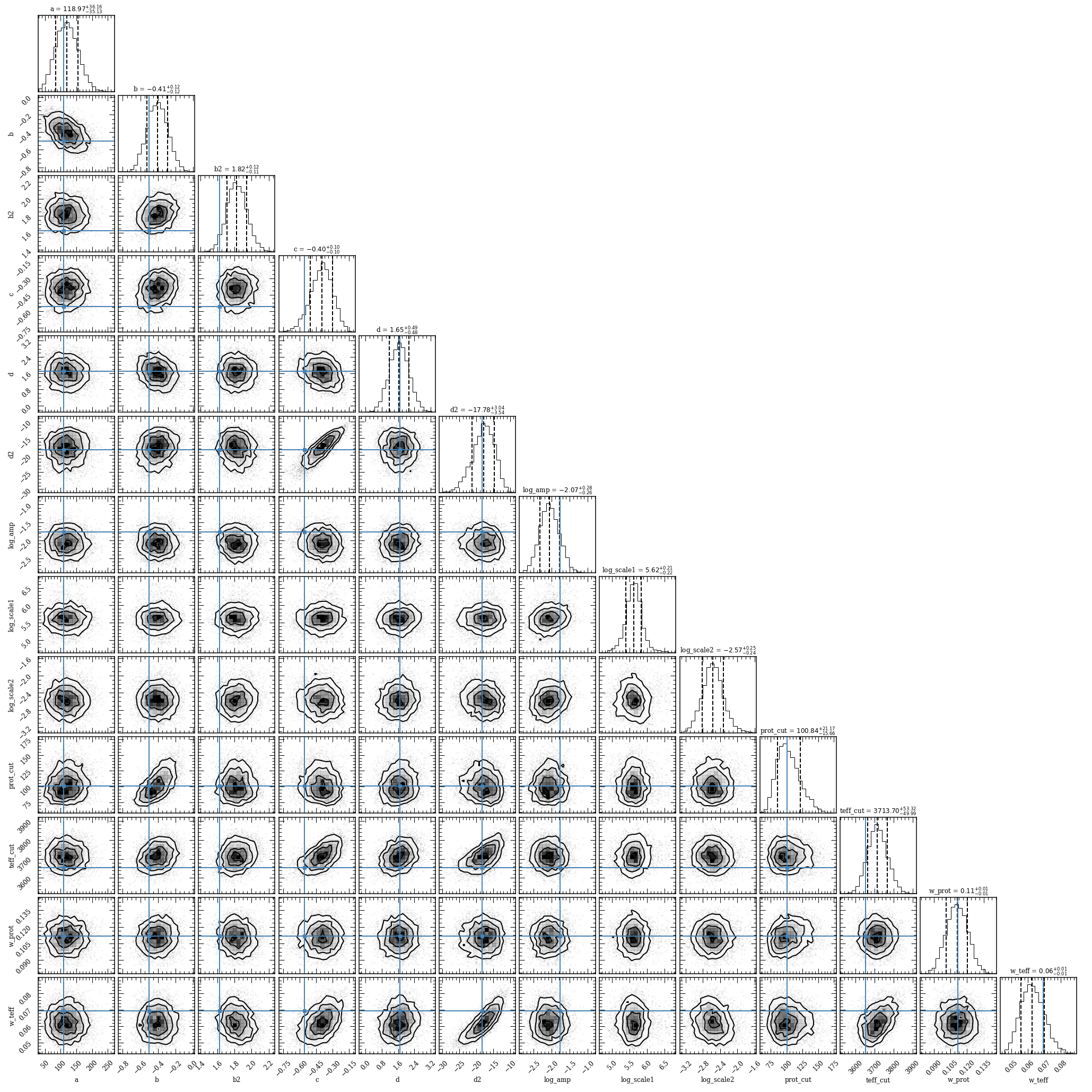}
    \caption{Parameter posterior distributions for the mean function of the Gaussian Process model for the partially convective stars after the MCMC has converged.
    The parameter descriptions are shown in Table~\ref{tab:tab1}. }
    \label{fig:corner}
\end{figure*}

\begin{figure*}[htb!]
    \centering
    \includegraphics[width=0.95\textwidth]{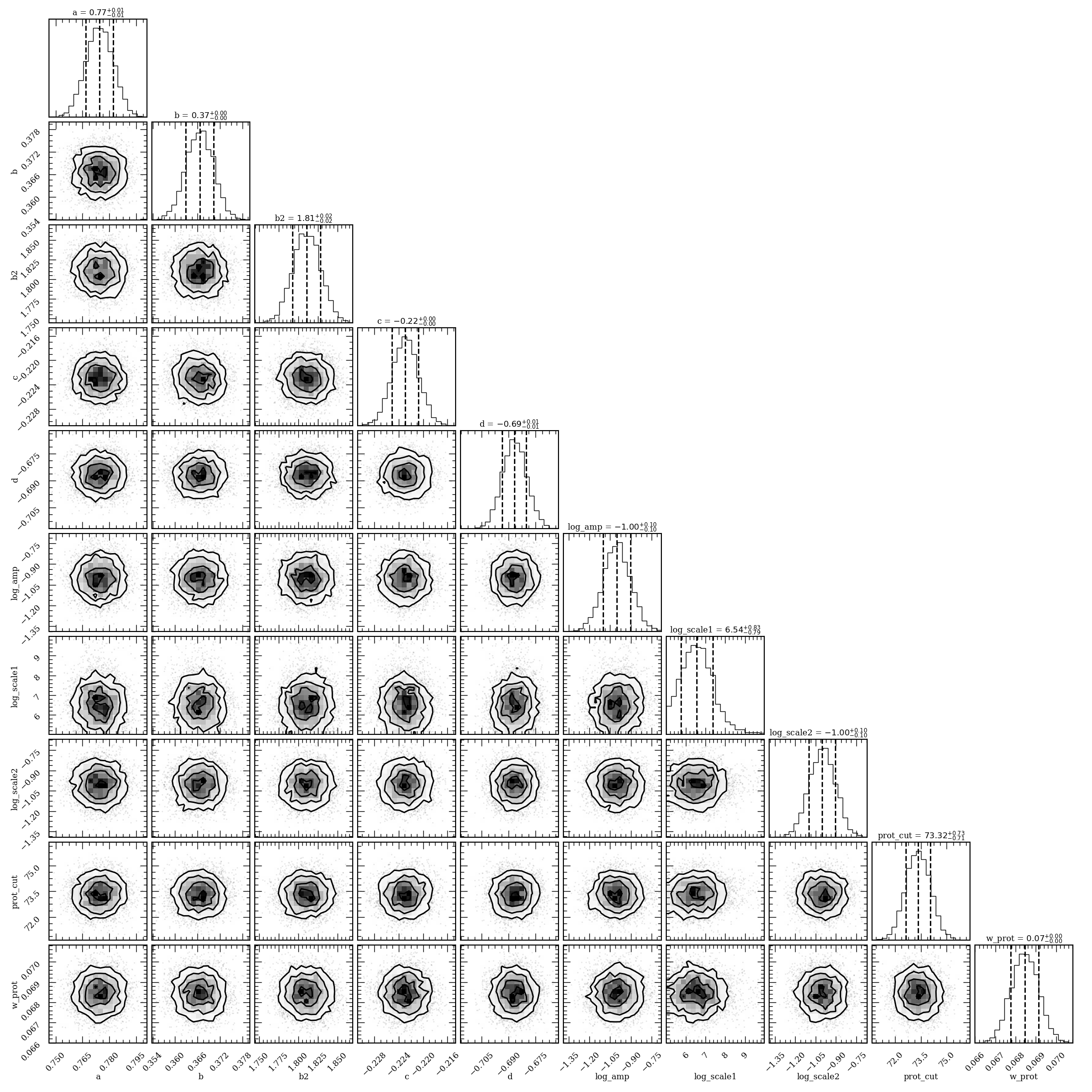}
    \caption{Same as Figure~\ref{fig:corner} but for the fully convective stars.}
    \label{fig:corner2}
\end{figure*}

We built the GP model using {\tt tinygp} \citep{tinygp}.
{\tt tinygp} is a {\tt PYTHON} library for building GP models.
It is built on {\tt jax} \citep{jax2018github}, which supports automatic differentiation that enables efficient model training.
We first optimized the parameters by maximizing the log-likelihood function, conditioned on the data described at the beginning of this section.
The optimized parameters were then used as initial inputs for the Markov chain Monte Carlo (MCMC) model to obtain the true distributions for the parameters.
The priors are Gaussians centered around the optimized parameters with a width described in Table~\ref{tab:tab1}. 
We implemented the MCMC model in {\tt numpyro} \citep{numpyro} for partially and fully convective stars separately.
The best-fit parameters for partially and fully convective stars are shown in Table~\ref{tab:tab1}, and the {\tt corner} \citep{corner} plots are shown in Figure~\ref{fig:corner} and Figure~\ref{fig:corner2} for partially and fully convective stars, respectively.

\subsubsection{Cross-validation}
\begin{figure}[hbt!]
    \centering
    \includegraphics[width=0.45\textwidth]{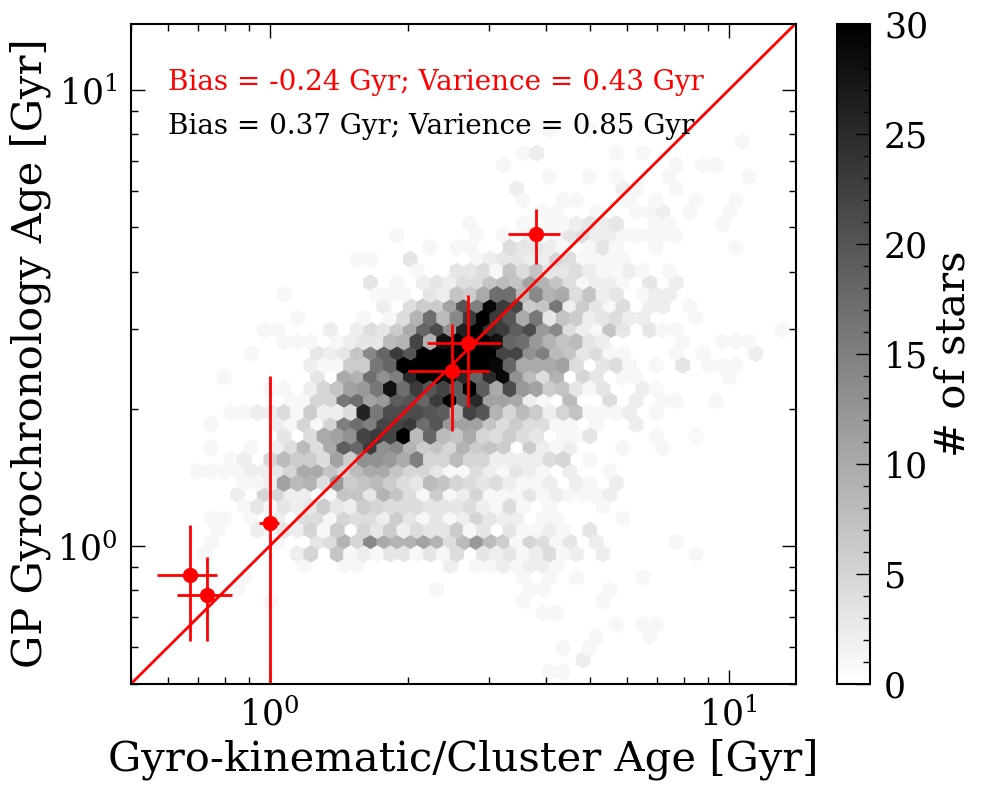}
    \caption{Cross-validation results for the 20\% gyro-kinematic age sample (black histogram) and individual clusters (red points).
    The systematic at 1 Gyr indicates existing bias in predicting stars younger than 1 Gyr old.}
    \label{fig:fig7}
\end{figure}

To ensure our model did not over-fit the data, we performed the cross-validation test by first excluding a random 20\% of the gyro-kinematic ages sample and optimized the model following the procedure described in the last section. 
The ages of these stars were then predicted using the trained model.
We also carried out a leave-one-out cross-validation test for the cluster sample by excluding a single cluster at a time, retraining the model, and predicting the age of that cluster with the trained model.
The cross-validation results are shown in Figure~\ref{fig:fig7}.
The $x$-error bars for the cluster sample are taken from the literature, and the $y$-error bars are calculated by taking the standard deviation of the predicted ages of all the cluster members. 
The average standard deviation ($y$-error bars) for the cluster cross-validation result is 0.62 Gyr.
The bias and variance for the cluster sample are -0.24 Gyr and 0.43 Gyr, respectively, and those for the gyro-kinematic ages are 0.37 Gyr and 0.85 Gyr, respectively. 
The cross-validation results suggest that our model is able to predict ages within $\sim$ 1 Gyr for main-sequence stars with reliable \prot, \bprp, and $M_G$ measurements. 
However, there exists a systematic at $\sim$1 Gyr in predicting gyro-kinematic ages, this systematic is most likely caused by the fact that the cluster sample between 0.67 Gyr to 1 Gyr occupies similar \prot-\teff\ space (see Figure~\ref{fig:fig3}), creating degeneracy in age predictions for stars younger than 1 Gyr. 
As a result, age predictions for stars $<$ 1 Gyr might be biased.
Stars around this age also occupies the \prot-\teff\ space where stars are expected to go through stalled spin-down \citep{Curtis2020}. 
Looking only at stars $>$ 5,000 K greatly reduces the pile-up. 

\section{Result} \label{sec:result}
\begin{figure*}
    \centering
    \includegraphics[width=0.98\textwidth]{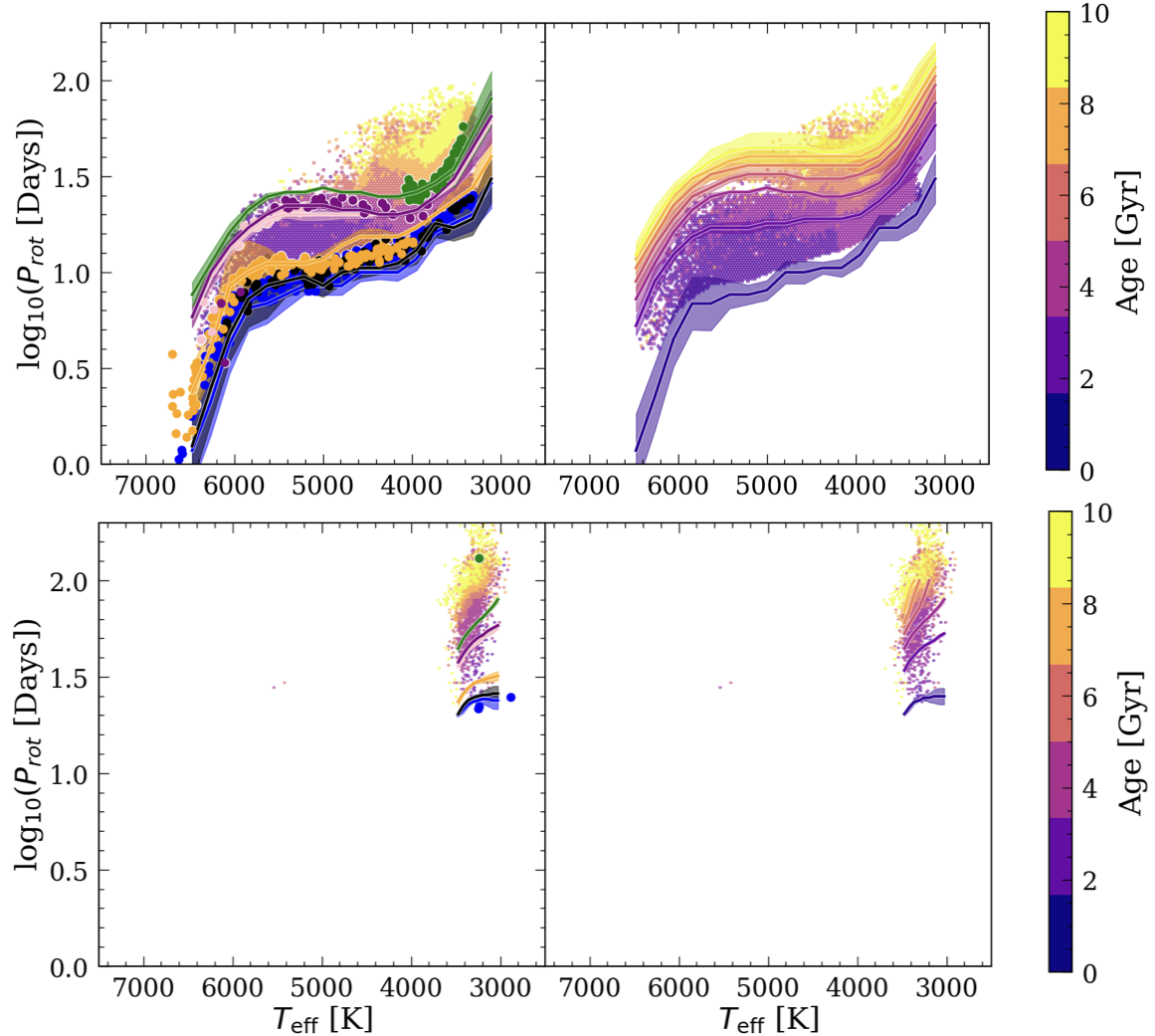}
    \caption{Running median (solid lines) and the standard deviation (shaded region) of 100 realizations of the GP models from this work for partially convective (top row) and fully convective (bottom row) stars.
    The models are overlaid on the full sample with gyro-kinematic ages. 
    The Jao's gap is used to distinguish between partially convective and fully convective stars. 
    {\it left column}: modeled isochrones (solid lines; the shaded area representing the model uncertainty) for each cluster (points).
    {\it right column}: Isochrones between 0.7 Gyr and 10 Gyr with a 1.55 Gyr separation colored by age. }
    \label{fig:fig6}
\end{figure*}

Figure~\ref{fig:fig6} shows the modeled isochrones for the cluster sample (left column) and the isochrones for 0.7 to 10 Gyr with 1.55 Gyr separations (right column) overlaid on the training sample of stars with gyro-kinematic ages that are partially convective (top row) and fully convective (bottom row). 
Since unlike most gyrochronology model, the model produced in this work infers ages from \prot\ and \teff\ instead of predicting rotation periods from age. 
As a result, constructing isochrone is not straightforward as we cannot input age as a direct input. 
The isochrones were calculated by first randomly drawing 100 model parameters from the MCMC fit and calculating the ages using these 100 models for the grid points in \teff-log$_{10}$(\prot) space, with the size of the grids to be [\teff, log$_{10}$(\prot)] = [52 K, log$_{10}$(1.1 Days)].
We then selected all (\prot, \teff) points that had predicted ages within 5\% of the desired age.
The running median (solid lines) and standard deviation (shaded area) of these grid points were finally calculated to be the model prediction and model uncertainty, respectively. 
Overall, our model traces the cluster sample well.
However, the model for fully convective stars cannot reproduce the one fully convective star in the open-cluster M67 (green point). 
This could be caused by the `edge effect' of the GP model or the gyro-kinematic ages used for training. 
In detail, since GPs cannot extrapolate, they tend adapt values that are close to the mean function outside of the range of the training data. 
Moreover, since obtaining gyro-kinematic ages requires binning stars in similar \prot, \teff, and $M_G$, they are less reliable at the edges because there are fewer stars in those bins. 
In addition, since fully convective stars could spin down faster than partially convective stars \citep[e.g.][]{Lu2023}, the bin size used to calculate gyro-kinematic ages could induce blurring as it will include stars of different ages.
Interestingly, there are stars with ages that match the M67 open-cluster age in the background gyro-kinematic age sample. 
This suggests some stars in this temperature and period range could be mis-classified as fully convective stars.
However, looking at these stars in the CMD, they are far away from the gap that is typically used to distinguish partial and fully convective stars \citep{Jao2018, vansanders2012}.
One other possibility is stars in that temperature and period range can have multiple ages.
Further study of the data and fully convective stars is needed to disentangle these scenarios.

\subsection{Predicting ages for the LEGACY dwarfs}
\begin{figure*}
    \centering
    \includegraphics[width=0.9\textwidth]{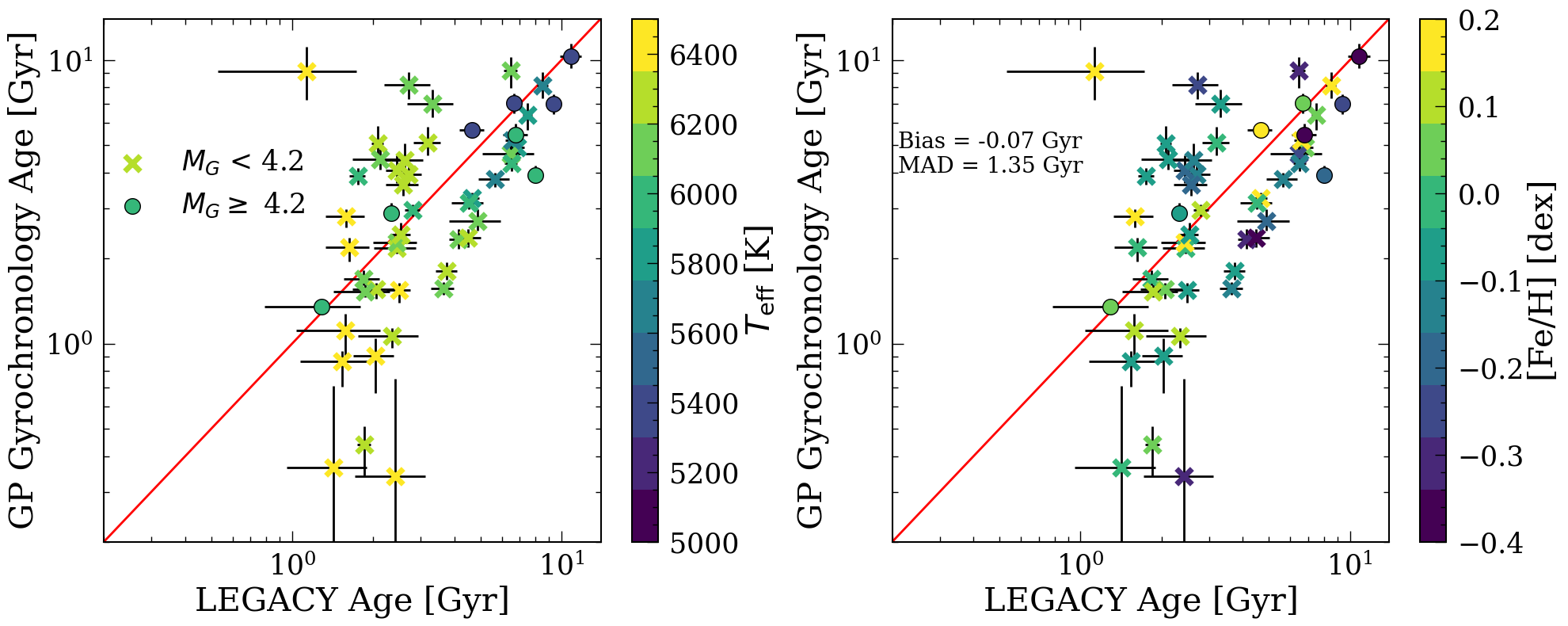}
    \caption{Testing result for 51 LEGACY stars with asteroseismic ages (not included in our training sample) colored by \teff\ (left; this work) and [Fe/H] \citep[right;][]{Silva2017}.
    Stars with $M_G <$ 4.2 mag (outside of the training sample) are shown in crosses. 
    This result suggests our model can estimate ages for single dwarf field stars with uncertainties just over 1 Gyr. }
    \label{fig:fig8}
\end{figure*}

To test our model, we predicted ages for 51 LEGACY dwarf stars with asteroseismic ages derived from Kepler \citep{Silva2017}, \prot, \teff, and $M_G$ data available from \cite{Santos2021}.
Figure~\ref{fig:fig8} shows the 1-to-1 comparison between the LEGACY asteroseismic ages and the gyrochronology ages from our model colored by \teff\ \citep[left;][]{Curtis2020} and [Fe/H] \citep[right;][]{Silva2017}. 
The uncertainties for the asteroseismic ages were calculated by taking the standard deviation of the age predictions from various pipelines from \cite{Silva2017}.
The ages and uncertainties for the gyrochronology ages were calculated by first calculating the ages for each star using 100 realizations of the model where the parameters are taken randomly from the MCMC fit. 
The 16th, 50th, and 84th percentile of the age predictions were then used to calculate the lower age limit, age, and higher age limit for each star.
The crosses show the stars with $M_G <$ 4.2 mag, which are outside of our training set. 
The bias and median absolute deviation (MAD) for the entire testing sample are -0.07 Gyr and 1.35 Gyr, respectively. 
This test suggests our model can estimate ages for single field dwarf stars with uncertainties just over 1 Gyr.

\begin{figure*}[hbt!]
    \centering
    \includegraphics[width=0.98\textwidth]{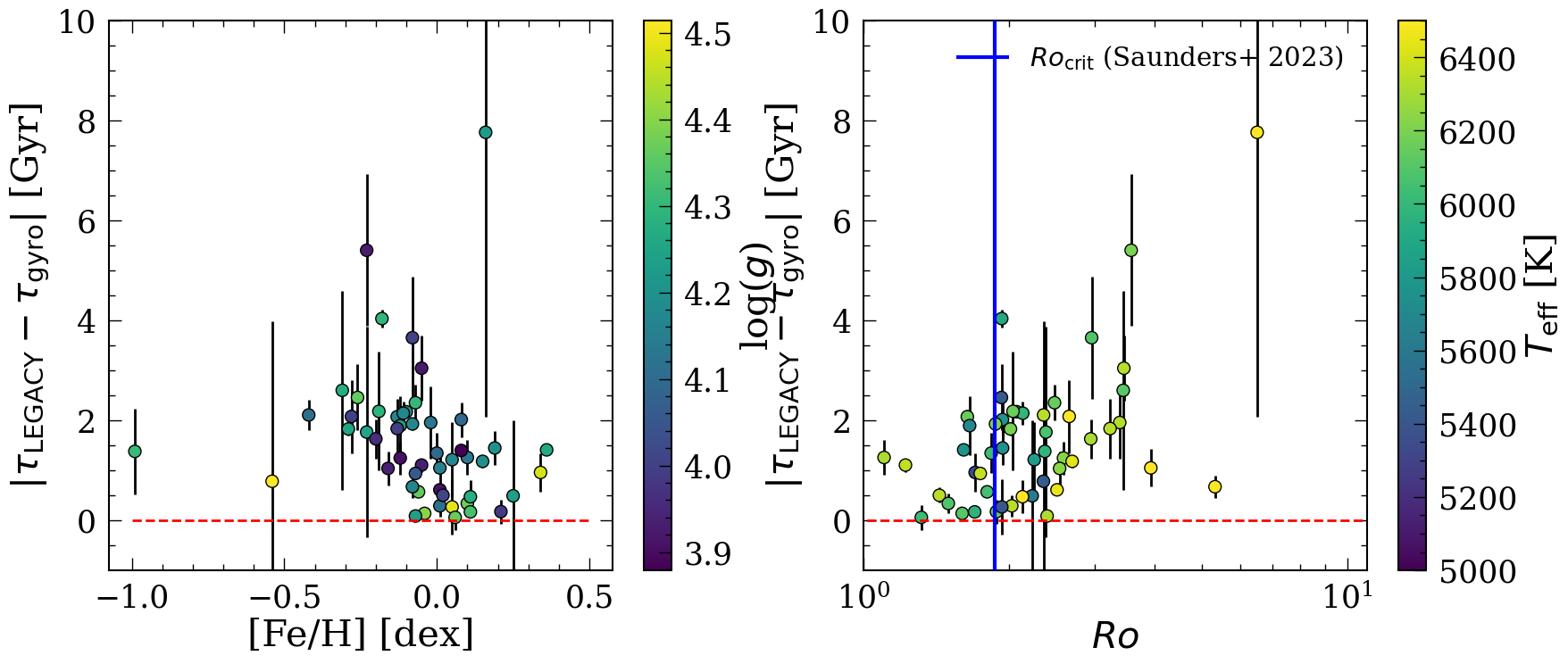}
    \caption{Absolute differences between the LEGACY ages and gyrochronology ages as a function of metallicity (left) and Rossby number (right).
    The red dotted lines show where the difference is 0.
    The uncertainties are calculated assuming Gaussian uncertainty ($\sigma^2=\sigma_{\rm LEGACY}^2+\sigma_{\rm gyro}^2$).
    There exists an obvious metallicity trend for stars with [Fe/H] $<$ 0.0 dex, in which the ages can deviate up to $\sim$ 2 Gyr as metallicity goes down to $\sim$ -0.5 dex.
    Age prediction significantly worsens for stars with Rossby number $> Ro_{\rm crit}$, which is $\sim$ 1.866 according to \cite{Saunders2023}.}
    \label{fig:feh}
\end{figure*}

Since our model did not take into account the effects of metallicity, we investigated this by plotting the absolute difference between the LEGACY and gyrochronology age against the metallicity of the star (Figure~\ref{fig:feh} left plot).
There is an obvious metallicity trend for stars with [Fe/H] $<$ 0.0 dex, suggesting future work of incorporating metallicity into this model is necessary \citep[also see Figure 9 in][for how metallicity can affect age determination using gyrochronology]{Claytor2020}. 
However, metallicity measurements that currently exist for low-mass stars are either limited in sample size or inaccurate and imprecise due to the presence of star spots and molecular lines in the spectra \citep[e.g.][]{Allard2011, Cao2022}.
As a result, we did not attempt to include training with metallicity in this work. 

As mentioned in the introduction, stars likely stop spinning down due to weakened magnetic braking after reaching a critical Rossby number, $Ro_{\rm crit}$ \citep[][]{vansaders2016}.
Recently, \cite{Saunders2023} fitted a magnetic braking model to asteroseismic and cluster data and concluded that $Ro_{\rm crit}$/$Ro_\odot$ = 0.91$\pm$0.03, which $Ro_{\rm crit}\sim$1.866 using \texttt{MESA} \citep{mesa}.
Indeed, the gyrochronology ages show large deviations from the asteroseismic ages for stars with \ro$>$1.866 (Figure~\ref{fig:feh} right plot). 
This suggests gyrochronology models should only be used to predict ages for stars with \ro$<$1.866.

\subsection{Gyrochronology Ages for $\sim$ 100,000 stars}
With this new gyrochronology relation, we predicted ages for $\sim$ 100,000 stars from Kepler \citep{McQuillan2014, Santos2021} and ZTF \citep[This work]{Lu2022} with period measurements, in which the ZTF periods were vetted using a random forest (RF) regressor trained on Gaia bp-rp color, absolute G magnitude, ruwe, and parallax. 
We did this by first training the RF on the ZTF periods that are highly vetted \citep{Lu2022}.
We then used the RF to predict the periods of the ZTF stars with measured periods described in section~\ref{subsec:periods}.
Finally, we selected period measurements that agree within 10\% of the predicted periods, which left us with 58,462 vetted ZTF periods with bp-rp color $>\sim$ 1.3 mag and period $>\sim$ 10 days.
We excluded stars with $Ro>$1.866, this left us with a final sample of 94,064 stars with periods from Kepler and ZTF.
We calculated the ages by using 100 realizations of the model with parameters taken randomly from the MCMC model after the chains had converged (same as what was done for the cluster isochrones in \prot-\teff~space and stars with asteroseismic ages).
We also tested how the measurement uncertainty in temperature and period could affect the ages by perturbing the measurements by 50 K and 10\%*$P_{\rm rot}$, respectively, assuming Gaussian errors.
We then recalculated the ages using these perturbed values.
We performed this 50 times for each star and found that the age uncertainty caused by the measurement error was negligible compared to the uncertainty in the model parameters.
Table~\ref{tab:catalog} shows the column description for this final catalog.

\begin{table}
\centering
\caption{Catalog description of the gyrochronology ages for $\sim$ 100,000 stars derived from this work.
This table is published in its entirety in a machine-readable format in the online journal.}
\begin{tabular}{crcrcr}
\hline
\hline
Column & Unit & Description\\
\hline
\texttt{source\_id} & & Gaia DR3 source ID\\
\texttt{KID}&  & Kepler input catalog ID if available \\
\texttt{Prot} & days & measured period \\
\texttt{bprp} & mag & de-redened \bprp\\
\texttt{abs\_G}& mag & absolute magnitude from Gaia DR3\\
\texttt{teff}& K & temperature derived from \texttt{bprp}\\
\texttt{Age}& Gyr & gyrochronology age \\
\texttt{Age\_err\_p}& Gyr & gyrochronology age upper uncertainty \\
\texttt{Age\_err\_m}& Gyr & gyrochronology age lower uncertainty \\
\hline
\end{tabular}
\label{tab:catalog}
\end{table}

\begin{figure}[hbt!]
    \centering
    \includegraphics[width=0.45\textwidth]{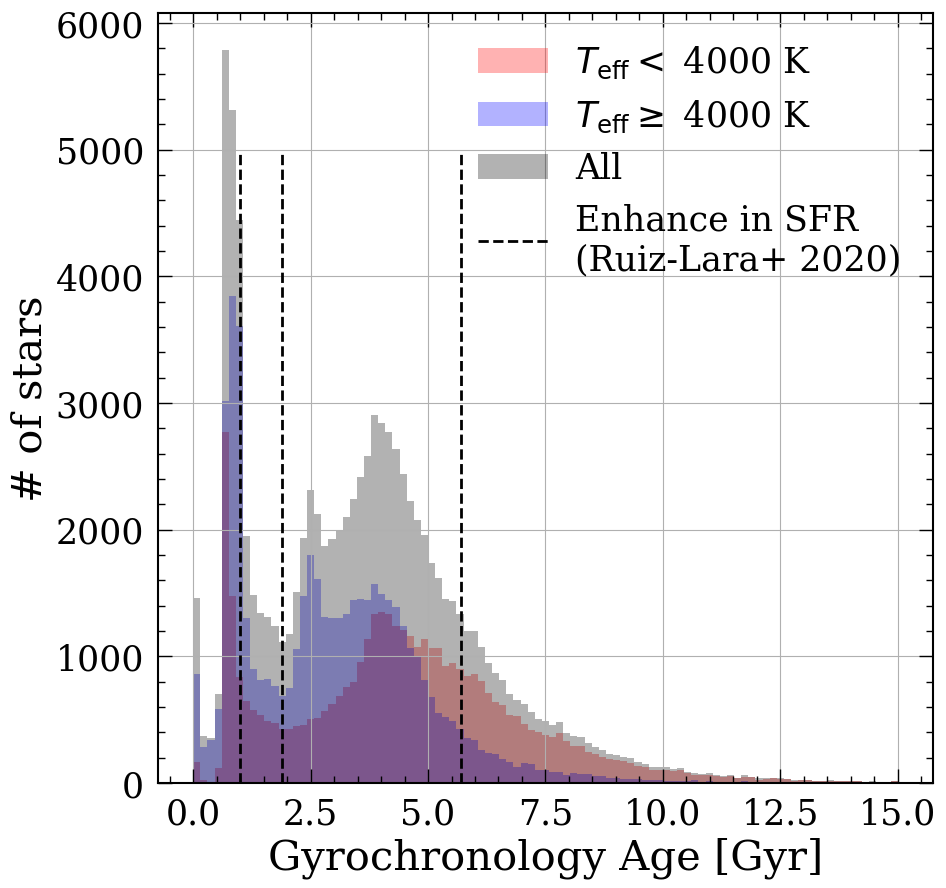}
    \caption{Histograms of stars with inferred gyrochronology ages from this work. 
    The black dotted lines show the recent enhancement of star formation rate (SFR) shown in \cite{ruizlara2020} (5.7, 1.9 and 1.0 Gyr).
    The peaks in the histograms can correspond to the enhancements of the star formation rate in the Milky Way (e.g., the peak in the tail around 5.7 Gyr), changes in stellar spin-down (e.g., the peak around 2.5 Gyr), or systematic bias (e.g., the peak around 1 Gyr).}
    \label{fig:fig_hist}
\end{figure}

Figure~\ref{fig:fig_hist} shows the histograms for stars with inferred gyrochronology ages using the calibrated relation from this work.
The black histogram shows the age distribution for the full sample of $\sim$ 100,000 stars, the red histogram shows those with \teff $<$ 4000 K, and the blue histogram shows those with \teff $\geq$ 4000 K. 
The black dotted lines show the recent enhancement of star formation rate (SFR) shown in \cite{ruizlara2020} (5.7, 1.9 and 1.0 Gyr).
The peaks in the histograms can correspond to the enhancements of the star formation rate in the Milky Way, changes in stellar spin-down, or systematic bias. 
For example, a peak exist in the tail of the distribution at the time of SFR enhancement 5.7 Gyr ago. 
This is the first time SFR enhancement has been shown using gyrochornology. 
However, some peak also correspond to limitations in the gyrochronology model.
For example, the peak around 2.5 Gyr exists only in stars $\geq$ 4000 K.
This peak most likely corresponds to the stall in spin-down for partially convective stars \citep[e.g.][]{Curtis2020} that do not exist for fully convective stars \citep[\teff $<$ 3500 K;][]{Lu2022}.
The stalling is thought to happen because the surface angular momentum loss is replenished by the core while the core and the envelope start re-coupling.
Depending on the re-coupling timescale, stars that span a range of ages will have very similar rotation period measurements, meaning they will have the same inferred age based on rotation and temperature alone. 
Future work should include other age indicators (e.g. stellar activity) to break this degeneracy. 

\subsection{Gyrochronology Ages for 384 Unique Planet Host Stars}
\begin{figure*}
\includegraphics[width=0.98\textwidth]{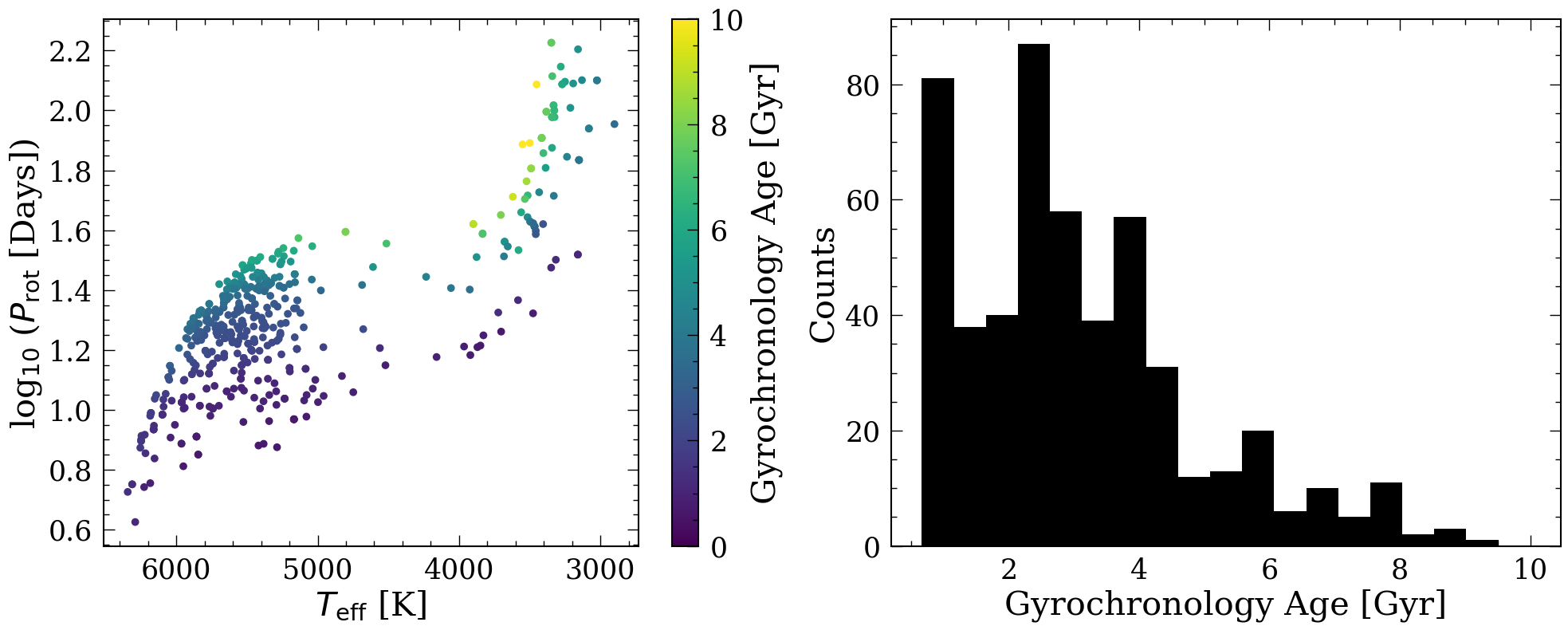}
    \caption{Left: \prot-\teff~diagram of exoplanet host stars colored by their gyrochronology ages inferred from this work. Right: histogram of the gyrochronology ages inferred from this work.}
    \label{fig:exo}
\end{figure*}

To infer gyrochronology ages for confirmed exoplanet host stars, we downloaded data from the NASA Exoplanet Archive\footnote{\url{https://exoplanetarchive.ipac.caltech.edu}}.
We combined stars with period measurements publicly available from the NASA Exoplanet Archive and from this work and inferred ages with 100 model realizations as done in the rest of this paper. 
We excluded stars with age prediction $<$ 0.67 Gyr and $Ro >$ 1.866, which left us with 384 unique planet host stars.
Within these stars, 338 have new rotation period measurements from \cite{Lu2022} and this work. 
Figure~\ref{fig:exo} shows the age distribution of these stars, and the column description for this catalog is shown in Table~\ref{tab:exo_cat}.

\begin{table*}
\centering
\caption{Catalog description of the gyrochronology ages for 384 exoplanet host stars derived from this work.
This table is published in its entirety in a machine-readable format in the online journal.}
\begin{tabular}{crcrcr}
\hline
\hline
Column & Unit & Description\\
\hline
\texttt{hostname} & & planet host name from the NASA Exoplanet Archive\\
\texttt{gaia\_id} & & Gaia DR2 source ID from the NASA Exoplanet Archive\\
\texttt{tic\_id}&  & TESS input catalog ID if available from the NASA Exoplanet Archive\\
\texttt{Prot} & days & measured period \\
\texttt{abs\_G}& mag & absolute magnitude from Gaia DR3\\
\texttt{teff}& K & temperature derived from \texttt{bprp}\\
\texttt{Age}& Gyr & gyrochronology age \\
\texttt{Age\_err\_p}& Gyr & gyrochronology age upper uncertainty \\
\texttt{Age\_err\_m}& Gyr & gyrochronology age lower uncertainty \\
\hline
\end{tabular}
\label{tab:exo_cat}
\end{table*}

\section{Limitations \& Future Work} \label{sec:limit}
Some possible limitations and biases of this model include:
\begin{itemize}
    \item {\it This model should only be applied to stars with $M_G >$ 4.2 mag, \prot\ $<$ 200 Days, ages $>$ 0.67 Gyr (or stars with \prot\ and \teff\ measurements above those of the members of the Praesepe), $Ro<$1.866, and 3,000 K $<$ \teff\ $<$ 7,000 K.}
    Inferring age for stars outside of this parameter space can lead to incorrect ages as the model is fully empirical, and stars with $Ro>$1.866 experienced weakened magnetic braking and stopped spinning down.
    However, Figure~\ref{fig:fig8} suggests the model still has strong predicting power for stars with $M_G >$ 3.5 mag. 
    \item {\it A systematic exist at $\sim$ 1 Gyr for stars $<$ 5,000 K.}
    The cluster sample suggests the isochrones for stars between 0.67 Gyr and 1 Gyr \citep[or even to 2.5 Gyr for low-mass stars due to stalling][]{Curtis2020} in \prot-\teff\ space have significant overlaps (see Figure~\ref{fig:fig2}), as a result, stars with a range of ages but similar \prot and \teff\ measurements will have similar age inference of around 1 Gyr. 
    \item {Inferring ages $\sim$ 2.5 Gyr for partially convective stars could be inaccurate.}
    Partially convective stars $\sim$ 2.5 Gyr start experiencing a stalled in their surface spin-down, most likely due to core-envelope decoupling \citep{Curtis2020}.
    As a result, stars with a range of ages can overlap in \prot-\teff\ space and create prediction biases at $\sim$ 2.5 Gyr.
    \item {\it No metallicity information is taken into account as reliable metallicity measurements for our sample are not yet available.} 
    Theory and observations strongly suggest a star with higher metallicity is likely to have a deeper convective zone and thus, spin-down faster \citep[e.g.][See et al.]{vanSaders2013,Karoff2018,Amard2019,Amard2020RotEvo}.
    As a result, strong bias can exist in age estimations using gyrochronology if assuming no metallicity variations exist in the sample \citep[e.g.][]{Claytor2020}.
    This means, all empirical gyrochornology relations available in the literature, calibrated on clusters or asteroseismic data, suffers from this bias.   
    Figure~\ref{fig:feh} shows the absolute differences between the LEGACY ages and gyrochronology ages ($\Delta$Age) as a function of metallicity.
    An obvious trend is observed that gyrochronology ages for lower metallicity stars deviate more from the asteroseismic ages.
\end{itemize}

\section{Conclusion} \label{sec:conc}
Gyrochronology is one of the few promising methods to age-date single main-sequence field stars.
However, gyrochronology relies strongly on empirical calibrations as the theories behind magnetic braking are complex and still unclear. 
The lack of a relatively complete sample of consistent and reliable ages for old, low-mass main-sequence stars with period measurements has prevented the use of gyrochronology for relatively old low-mass stars beyond $\sim$ 4 Gyr \citep[the age of the oldest cluster with significant period measurements][]{Dungee2022}. 

By combining period measurements from Kepler and ZTF, using the gyro-kinematic age-dating method, we constructed a large sample of reliable kinematic ages expanding the \prot-\teff\ space that is most suitable for gyrochronology (4 days $<$ \prot $<$ 200 days; 3,000 K $<$ \teff $<$ 7,000 K).
By using a Gaussian Process model, we constructed the first calibrated gyrochronology relation that extends to the fully convective limit and is suitable for stars with ages between 0.67 Gyr and 14 Gyr. 
Cross-validation tests and predicting ages for dwarf stars with asteroseismic signals suggest our model can provide reliable ages with uncertainties on the order of 1 Gyr, similar to that of isochrone ages \citep[e.g.][Figure 9]{Berger2023}.
In this paper, we provide ages for $\sim$ 100,000 stars with period measurements from Kepler and ZTF, of which 763 are exoplanet host stars with a total of 1,060 planets.

Systematic exist at stellar age $\sim$ 1 (for \teff$<$ 5,000 K) and 2.5 Gyr (for partially convective stars) due to the fact that stars with a range of ages overlap in \teff-\prot\ space, most likely due to stalling caused by core-envelope decoupling.
This causes the model to infer similar ages for stars of a range of ages. 
Adding other age indicators such as stellar activity in the future could potentially break the degeneracy in \teff-\prot\ space for stars of certain range of ages.
Obvious metallicity bias exists for this model (see Figure~\ref{fig:feh} left plot; deviation of $\sim$2 Gyr from the asteroseismic ages as the metallicity of the star reaches -0.5 dex), as a result, future work should incorporate metallicity measurements.

\section{acknowledgments}
Y.L would like to thank Joel Ong for suggesting the title. 
R.A. acknowledges support by NASA under award \#80NSSC21K0636 and NSF under award \#2108251.
This work has made use of data from the European Space Agency (ESA)
mission Gaia,\footnote{\url{https://www.cosmos.esa.int/gaia}} processed by
the Gaia Data Processing and Analysis Consortium (DPAC).\footnote{\url{https://www.cosmos.esa.int/web/gaia/dpac/consortium}} Funding
for the DPAC has been provided by national institutions, in particular
the institutions participating in the Gaia Multilateral Agreement.
This research also made use of public auxiliary data provided by ESA/Gaia/DPAC/CU5 and prepared by Carine Babusiaux.

This research has made use of the NASA Exoplanet Archive, which is operated by the California Institute of Technology, under contract with the National Aeronautics and Space Administration under the Exoplanet Exploration Program.

This research was done using services provided by the OSG Consortium \citep{OSG1,OSG2}, which is supported by the National Science Foundation awards \#2030508 and \#1836650.

This research has also made use of NASA's Astrophysics Data System, 
and the VizieR \citep{vizier} and SIMBAD \citep{simbad} databases, 
operated at CDS, Strasbourg, France.

%

\vspace{5mm}
\facilities{Gaia, Kepler, TESS, PO:1.2m (ZTF), Exoplanet Archive}


\software{  \texttt{Astropy} \citep{astropy:2013, astropy:2018},
             \texttt{dustmaps} \citep{Green2018},
            \texttt{Matplotlib} \citep{matplotlib}, 
            \texttt{NumPy} \citep{Numpy}, 
            \texttt{Pandas} \citep{pandas}, 
            \texttt{tinygp}
            \citep{tinygp},
            \texttt{numpyro}
            \citep{numpyro}
            }



\appendix
\section{Visualizing the gyrochronology model}
\begin{figure*}[hbt!]
    \centering
    \includegraphics[width=0.98\textwidth]{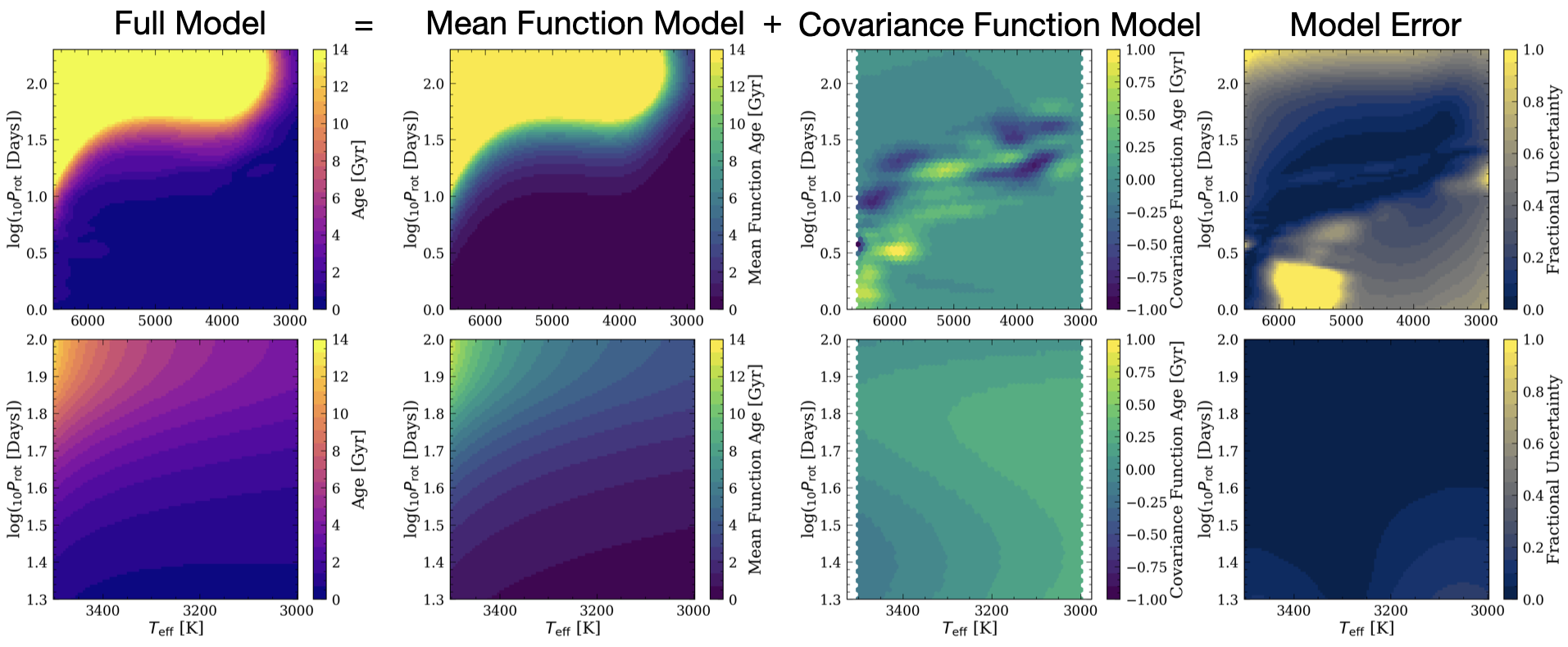}
    \caption{Full model in the training parameter space (left column), the mean function prediction (second column), and the covariance function correction (third column) for partially convective (top row) and fully convective (bottom row) stars.
    The last column shows the age uncertainty associated with the model parameters. }
    \label{fig:fig5}
\end{figure*}

We can visualize the gyrochronology model by looking at the best-fit mean and covariance functions separately. 
Figure~\ref{fig:fig5} shows our full model in the training parameter space (left column), the mean function prediction (second column), and the covariance function correction (third column) for partially convective (top row) and fully convective (bottom row) stars.
The last column shows the age uncertainty associated with the model parameters. 
The uncertainty is calculated based on 100 realization of the model with parameter drown from the MCMC posterior distribution. 
The large fractional uncertainty for partically convective stars around 6,000 K is both caused by the young age and the overlapping isochrones in the cluster training data (see Figure~\ref{fig:fig6}).

\bibliography{sample631}{}
\bibliographystyle{aasjournal}



\end{document}